\documentclass[letterpaper,prd,10pt,showpacs,showkeys,preprintnumbers,superscriptaddress,nofootinbib,preprint]{revtex4-1}
\usepackage{amsmath,amssymb,amsfonts,dsfont,mathrsfs,amsthm,mathtools,graphicx}
\usepackage{color}
\usepackage[usenames]{xcolor}
\usepackage{hyperref}
\usepackage{siunitx}
\usepackage{bigints}
\usepackage[english]{babel}
\hypersetup{linktocpage,colorlinks=true,urlcolor=blue,linkcolor=blue,citecolor=blue}
\usepackage{float}
\usepackage[paperwidth=621pt,top=65pt,right=60pt,bottom=80pt,left=60pt,headheight=50pt]{geometry}
\usepackage[normalem]{ulem}

\newcommand{\diff}[1]{\text{d}#1}

\begin{document}

\title{Electric/magnetic duality of dyonic Kerr-Newman-NUT-AdS spacetimes}

\author{Crist\'obal \surname{Corral}}
\email{cristobal.corral@uai.cl}
\affiliation{Departamento de Ciencias, Facultad de Artes Liberales, Universidad Adolfo Ibáñez, Avenida Padre Hurtado 750, 2562340, Viña del Mar, Chile}

\author{Rodrigo \surname{Olea}}
\email{rodrigo\_olea\_a@yahoo.co.uk}
\affiliation{Instituto de F\'isica, Pontificia Universidad Cat\'olica de Valpara\'iso, Casilla 4059, Valpara\'iso, Chile}

\begin{abstract}
We study the (anti-)self-duality conditions under which the electric and magnetic parts of the conserved charges of the dyonic Kerr-Newman-NUT-AdS solution become equivalent. Within a holographic framework, the stress tensor and the boundary Cotton tensor are computed from the electric/magnetic content of the Weyl tensor. The holographic stress tensor/Cotton tensor duality is recovered along the (anti-)self-dual curve in parameter space. We show that the latter not only implies a duality relation for the mass but also for the angular momentum.  The partition function is computed to first order in the saddle-point approximation and a BPS bound is obtained. The ground state of the theory is enlarged to all the (anti-)self-dual configurations when the $SO(4)$ and $U(1)$ Pontryagin densities are introduced. We demonstrate this at the level of the action and variations thereof.  
\end{abstract}

\maketitle

\section{Introduction}

Asymptotically anti-de Sitter (AdS) black holes are excellent theoretical laboratories for testing the AdS/CFT correspondence~\cite{Maldacena:1997re,Gubser:1998bc,Witten:1998qj}. One of the most remarkable examples is the description of confinement/deconfinement phases in $\mathcal{N}=4$ super Yang-Mills theory by means of the Hawking-Page transition of Schwarzschild-AdS black hole in the bulk theory~\cite{Witten:1998zw}. The thermodynamic phase space of planar Reissner-Nordstr\"om-AdS black hole, on the other hand, provides a holographic dual to catastrophe theory~\cite{Chamblin:1999tk}. Holographic superconductors, in turn, can be modeled by asymptotically AdS black holes with scalar hair~\cite{Hartnoll:2008vx,Hartnoll:2008kx} and different response functions have been computed in Refs.~\cite{Albash:2008eh,Horowitz:2010gk,Herzog:2010vz,Montull:2012fy,Lala:2014jca}. Indeed, asymptotically AdS black holes with planar transverse sections can be used to describe holographic systems with momentum dissipation if scalar fields with axionic profiles are considered~\cite{Andrade:2013gsa,Andrade:2014xca}. The Kerr-AdS black hole also offers an interesting scenario for holographic applications, ranging from rotating holographic superconductors~\cite{Sonner:2009fk}, anisotropic drag force of heavy quarks~\cite{NataAtmaja:2010hd}, rotational holographic transport~\cite{Meert:2024dud}, to warped-AdS/CFT when its extreme version is considered~\cite{Guica:2008mu,Compere:2012jk}. 

Vorticity in the dual CFT can be obtained by analyzing the holographic stress-energy tensor of the Taub-NUT-AdS metric~\cite{Leigh:2011au,Caldarelli:2011idw,Leigh:2012jv,Kalamakis:2020aaj}. This spacetime was first found in Refs.~\cite{Taub:1950ez,Newman:1963yy} and it endows the space with a magnetic source of mass---the NUT charge---and a topological defect dubbed the Misner string~\cite{Misner:1963fr}. The Taub-NUT metric is usually regarded as the gravitational analog of an electromagnetic dyon~\cite{Lynden-Bell:1996dpw,Nouri-Zonoz:1997mnd,Nouri-Zonoz:1998whb} since the dominant contributions of the electric and magnetic parts of the Weyl tensor are captured by the Ashtekar-Magnon-Das (AMD) charges~\cite{Ashtekar:1984zz,Ashtekar:1999jx}. Indeed, at the (anti-)self-dual point of this space, a remarkable duality between its conserved charges occurs, which can be translated into a holographic stress tensor/Cotton tensor duality~\cite{Leigh:2007wf,Mansi:2008br,Mansi:2008bs,Bakas:2008gz,Bakas:2009pbm,Miskovic:2009bm,Caldarelli:2011idw,Leigh:2011au,Leigh:2012jv,Mukhopadhyay:2013gja,Araneda:2016iiy,Araneda:2018orn}. Here, we explore a similar type of duality in the presence of rotation and Maxwell fields: this is found in Kerr-Newman-NUT-AdS (KNNAdS) black hole and it will be the main subject of this work.   

The dyonic Kerr-Newman-NUT-AdS spacetime is a five-parameter family of electrovacuum solutions that belongs to the Plebanski-Demianski family, being Petrov type D~\cite{Plebanski:1976gy}. These spaces possess a hidden symmetry related to the existence of a Killing-Yano tensor~\cite{Demianski:1980mgt}. Recently, it has been shown that the latter is associated with the existence of a hidden two-dimensional conformal symmetry and its extremal limit resembles similar properties as the extremal Kerr black hole, allowing one to compute the black hole entropy from a Cardy-like formula~\cite{Sakti:2017pmt,Sakti:2020jpo}. The higher-dimensional generalization of the Kerr-Newman-NUT-AdS black hole has been reported in Ref.~\cite{Chen:2006xh} and their hidden symmetries were discussed in Ref.~\cite{Kubiznak:2006kt}. Additionally, the integrability of geodesic motion in arbitrary dimensions has been studied in Ref.~\cite{Page:2006ka} and the separability of Hamilton-Jacobi and Klein-Gordon equations on this background was demonstrated in Refs.~\cite{Dadhich:2001sz,Frolov:2006pe}. In the absence of Maxwell fields, its thermodynamics, alongside a generalized Smarr relation and their Brown-York charges have been obtained in Ref.~\cite{Rodriguez:2021hks}. Supersymmetric properties of this solution have been studied in Ref.~\cite{Alonso-Alberca:2000zeh}, showing that they preserve generically $1/4$ of the total supersymmetry.

The analytic continuation of the dyonic Kerr-Newman-NUT-AdS black hole possesses a curve in the parameter space that renders both the Weyl tensor and the $U(1)$ field strength (anti-)self-dual~\cite{Alonso-Alberca:2000zeh}. On this curve, we show that the electric and magnetic AMD charges are equal and there appears a nontrivial relation between the mass and the NUT charge. Remarkably, this relation not only establishes an electric/magnetic duality at the level of the black hole's mass but also at the level of the angular momentum. This, in turn, leads to a nontrivial relation between the holographic stress tensor and the boundary Cotton tensor~\cite{Leigh:2007wf,Caldarelli:2011idw,Mansi:2008br,Mansi:2008bs,Miskovic:2009bm,Bakas:2008gz,Bakas:2009pbm,Leigh:2011au,Leigh:2012jv,Mukhopadhyay:2013gja}, allowing us to compute these quantities explicitly (see Eq.~\eqref{selfdualtaucotton} below). This relation also holds in the absence of the rotation parameter~\cite{Ciambelli:2020qny}; however, an arbitrary position of the Misner string---the gravitational analog of the Dirac string---is required. The partition function and Noether-Wald charges are computed using topological renormalization; the former is obtained to first order in the saddle-point approximation. In four dimensions, this renormalization scheme coincides with holographic renormalization~\cite{Balasubramanian:1999re,Emparan:1999pm,deHaro:2000vlm}, with the advantage of providing a covariant prescription for computing conserved charges in vacuum Einstein-AdS gravity~\cite{Anastasiou:2020zwc}. We show that this prescription not only works for Einstein-AdS gravity but also in the presence of Maxwell fields due to its falloff dictated by conformal symmetry, similar to the case of Einstein-AdS-Maxwell gravity with intrinsic counterterms~\cite{Chamblin:1999tk,Caldarelli:1999xj}. Even more, along the (anti-)self-dual curve in parameter space, the renormalized Euclidean on-shell action can be written as a linear combination of the $SO(4)$ and $U(1)$ Chern-Pontryagin indices, saturating a Bogomol'nyi–Prasad–Sommerfield (BPS) bound. We show that the addition of topological terms of the Pontryagin class enlarges the ground state of the theory to all (anti-)self-dual configurations of the Einstein-AdS-Maxwell theory. This is shown at the level of the action and variations thereof.

The article is organized as follows: In Sec.~\ref{sec:KNNAdS}, we review the Kerr-Newman-NUT-AdS black hole and fix notation. Section~\ref{sec:topologicalren} studies topological renormalization of Einstein-AdS-Maxwell theory. Then, in Sec.~\ref{sec:SD}, we study the conditions under which the Kerr-Newman-NUT-AdS solution becomes (anti-)self-dual and show how the inclusion of Chern-Pontryagin terms allows one to enlarge the ground state of the theory. In Sec.~\ref{sec:AMD}, we compute asymptotic charges and obtain the holographic stress tensor explicitly. Finally, in Sec.~\ref{sec:discussion} we summarize and present our final remarks.

\section{The Kerr-Newman-NUT-AdS black hole\label{sec:KNNAdS}}

The Kerr-Newman-NUT-AdS black hole is a five-parameter family of solutions of the Einstein-Maxwell system with negative cosmological constant. It was first found by Carter~\cite{Carter:1968ks} and their global properties were studied by Miller in Ref.~\cite{Miller:1973hqu}. Gibbons and Perry constructed a gravitational instanton from its analytic continuation in the absence of Maxwell fields~\cite{Gibbons:1979nf}. Then, a generalization of this solution was found by Plebanski and Demianski by including acceleration~\cite{Plebanski:1976gy}; this six-parameter family of solutions contains the Kerr-Newman-NUT-AdS black hole as a particular case (for a more recent discussion see Ref.~\cite{Griffiths:2005qp}). The anz\"atze describing the line element and $U(1)$ potential of this spacetime can be parametrized as
\begin{subequations}\label{ansatz}
    \begin{align}\notag
    \diff{s^2} &= - \frac{\Delta_r(r)}{\Xi^2\Sigma(r,\vartheta)}\left[\diff{t} + \left(2n\cos\vartheta+2nC-a\sin^2\vartheta\right)\diff{\varphi} \right]^2 + \frac{\Sigma(r,\vartheta)}{\Delta_r(r)}\diff{r^2} \\
    &\quad + \frac{\Sigma(r,\vartheta)}{\Delta_\vartheta(\vartheta)}\diff{\vartheta^2} + \frac{\Delta_\vartheta(\vartheta)\sin^2\vartheta}{\Xi^2\Sigma(r,\vartheta)}\left[a\diff{t}-\left(r^2+a^2+n^2-2anC \right)\diff{\varphi} \right]^2\,, \\
    A &= A_\mu \diff{x^\mu} = a_t(r,\vartheta)\diff{t} + a_\varphi(r,\vartheta)\diff{\varphi}\,,
\end{align}
\end{subequations}
respectively, where $a$ is a rotation parameter, $\Xi = 1 - a^2/\ell^2$, the NUT charge is denoted by $n$, and $C=\pm1,0$ is a discrete parameter that controls the position of the Misner string. This is a stationary and axisymmetric spacetime that remains invariant under the flow of two Killing vectors, i.e. $\partial_t$ and $\partial_\varphi$. Notice that we have rescaled the time coordinate such that the algebra of conserved charges of the solution~\eqref{solution} is $\mathfrak{so}(2,3)$ (see Refs.~\cite{Kostelecky:1995ei,Rodriguez:2021hks}).

The Einstein-AdS-Maxwell system is described by the field equations
\begin{subequations}\label{eom}
\begin{align}\label{eomg}
   R_{\mu\nu} - \frac{1}{2}g_{\mu\nu}R -\frac{3}{\ell^2} g_{\mu\nu} &= \frac{1}{2\kappa} T_{\mu\nu}\,, \\
    \label{eomA}
    \nabla_\mu F^{\mu\nu} &= 0\,,
\end{align}    
\end{subequations}
where $R_{\mu\nu}=R^{\lambda}_{\ \mu\lambda\nu}$ and $R=g^{\mu\nu}R_{\mu\nu}$ are the Ricci tensor and scalar, respectively, $\ell$ is the AdS radius, the gravitational constant is denoted by $\kappa=(16\pi G)^{-1}$, the Maxwell's field strength is $F_{\mu\nu}=\partial_\mu A_\nu - \partial_\nu A_\mu$, while its stress-energy tensor can be expressed as
\begin{align}\label{Tmunu}
    T_{\mu\nu} = F_{\mu\lambda}F_{\nu}^{\ \lambda} - \frac{1}{4}g_{\mu\nu}F_{\lambda\rho}F^{\lambda\rho}\,.
\end{align}
In four dimensions, the trace of the stress-energy tensor for the Maxwell fields vanishes identically as a consequence of its conformal invariance. 

The Kerr-Newman-NUT-AdS spacetime is a solution of the Einstein-AdS-Maxwell field equations~\eqref{eom} whose metric functions and Maxwell potential are given by 
\begin{subequations}\label{solution}
    \begin{align}
    \Delta_r(r) &= r^2 - 2mGr + \frac{q^2+p^2}{4\kappa} + \frac{1}{\ell^2}\left[r^2(r^2+6n^2+a^2)+(3n^2+\ell^2)(a^2-n^2) \right]\,, \\
    \Sigma(r,\vartheta) &= r^2 + (n+a\cos\vartheta)^2\,,\\
    \Delta_\vartheta(\vartheta) &= 1 - \frac{a\cos\vartheta(4n+a\cos\vartheta)}{\ell^2}\,, \\
    a_t(r,\vartheta) &= - \frac{qr+p(n+a\cos\vartheta)}{\Xi\Sigma(r,\vartheta)}\,, \\
    a_\varphi(r,\vartheta) &= \frac{qr\left[a^2+n^2-2anC-(n+a\cos\vartheta)^2\right]}{a\Xi\Sigma(r,\vartheta)} + \frac{p\left(n+a\cos\vartheta \right)\left(r^2+a^2+n^2-2anC\right)}{a\Xi\Sigma(r,\vartheta)}\,,
\end{align}
\end{subequations}
where $m$, $a$, $q$, and $p$ are integration constants related to the mass, rotation, electric, and magnetic charges, respectively. In the presence of the NUT charge, the ring singularity of the Kerr-AdS metric is removed if and only if $|n|\neq a$, although the solution could have conical singularities~\cite{Mukherjee:2018dmm}. This can be seen directly from the Kretschmann invariant
\begin{align}\label{kret}
    R_{\mu\nu\lambda\rho}R^{\mu\nu\lambda\rho} = \frac{H(r,\vartheta)}{\left[r^2+(n+a\cos\vartheta)^2\right]^6}\,,
\end{align}
where $H(r,\vartheta)$ is monotonically increasing function of the radial coordinate $\forall\, \vartheta\in [0,\pi]$ and it behaves as $H(r,\vartheta)\sim r^{12}$ as $r\to\infty$. Thus, one can see that the Kretschmann invariant becomes asymptotically constant and, as $r\to0$, it remains finite $\forall (r,\vartheta)$ as long as $|n|\neq a$. 

The solution has an event horizon located at the locus $r=r_+$, defined as the largest positive root of $\Delta(r_+)=0$. The Killing vector field that becomes null at the horizon is
\begin{align}\label{KillingHorizon}
    \xi = \partial_t + \Omega_+\partial_\varphi \,,\;\;\;\;\; \mbox{where} \;\;\;\;\; \Omega_+ = \frac{a}{r_+^2+a^2+n^2-2anC} \,,
\end{align}
denotes the angular velocity of the horizon. It is well known that rotating black holes in AdS have a nonvanishing angular velocity at infinity~\cite{Caldarelli:1999xj,Gibbons:2004ai}. When the NUT charge is switched on, the asymptotic angular velocity is not even constant~\cite{Durka:2019ajz}. Nevertheless, it is still possible to obtain a first law of thermodynamics and a Smarr relation as shown in Ref.~\cite{Rodriguez:2021hks} and references therein (see also~\cite{Bordo:2019tyh} in the absence of rotation). The mass parameter is related to the horizon radius of the black hole and its electric and magnetic charges through 
\begin{align}
    m &= \frac{1}{2G\ell}\left[r_+^3 + \left(a^2+\ell^2+6n^2\right)r_+ + \frac{\left( \ell^2+3n^2\right)\left(a^2-\ell^2\right)}{r_+}\right] + \frac{2\pi\left(q^2+p^2 \right)}{r_+}\,.
\end{align}
The Hawking temperature of the solution is $T_H=\tfrac{\kappa_s}{2\pi}$, where $\kappa_s$ is the surface gravity generated by the Killing vector of Eq.~\eqref{KillingHorizon}. Its value is explicitly given by
\begin{align}\label{TH}
    T_H = \frac{\Omega_+}{4\pi a r_+\Xi}\left\{4\pi G\left(q^2+p^2 \right) - \left(r_+^2+n^2-a^2 \right) - \frac{1}{\ell^2}\left[3\left(r_+^2+n^2 \right)^2 + a^2\left(r_+^2-3n^2 \right)\right] \right\}\,.
\end{align}
As $T_H\to0$, the solution becomes extremal and the near-horizon geometry has an enhanced $SL(2,\mathbb{R})\times U(1)$ isometry group. In Ref.~\cite{Sakti:2020jpo}, the authors computed the microscopic black hole entropy from a Cardy-like formula, recovering the Bekenstein-Hawking entropy (see also Ref.~\cite{Perry:2022udk}). 

In the next section, we show that the Gauss-Bonnet term can be used to renormalize the Einstein-AdS-Maxwell action. This is a well-established result at the level of vacuum solutions~\cite{Aros:1999id,Aros:1999kt,Olea:2005gb}. Also, employing techniques in the framework of gauge/gravity duality, it was proved to be equivalent to holographic renormalization~\cite{,Miskovic:2009bm,Anastasiou:2020zwc} and also naturally leads to the Conformal Mass definition \cite{Jatkar:2014npa}. In the following, we extend this scheme in the presence of vector couplings.

\section{Topological renormalization of Einstein-AdS-Maxwell \label{sec:topologicalren}}

The bulk dynamics of the four-dimensional Einstein-AdS-Maxwell theory is given by the field equations~\eqref{eom}, that can be obtained by performing arbitrary variations of the action principle
\begin{align}\label{EMaction}
    I_{\rm EM}[g_{\mu\nu},A] = \kappa\int_{\mathcal{M}}\diff{^4x}\sqrt{|g|}\left(R+ \frac{6}{\ell^2} + \alpha\mathcal{G} \right) - \frac{1}{4}\int_{\mathcal{M}}\diff{^4x}\sqrt{|g|}\,F_{\mu\nu}F^{\mu\nu}\,.
\end{align}
Here, $\alpha$ is the coupling constant of the Gauss-Bonnet term; the latter is defined as
\begin{align}\label{GB}
    \mathcal{G}  = \frac{1}{4}\delta^{\mu_1\ldots\mu_4}_{\nu_1\ldots\nu_4}R^{\nu_1\nu_2}_{\mu_1\mu_2}R^{\nu_3\nu_4}_{\mu_3\mu_4} = R^2 - 4R_{\mu\nu}R^{\mu\nu} + R_{\mu\nu\lambda\rho}R^{\mu\nu\lambda\rho}\,,
\end{align}
with $\delta^{\mu_1\ldots\mu_p}_{\nu_1\ldots\nu_p}=p!\,\delta^{[\mu_1}_{[\nu_1}\dots\delta^{\mu_p]}_{\nu_p]}$ being the generalized Kronecker delta. The inclusion of the Gauss-Bonnet term does not modify the bulk dynamics since it can be written as a boundary integral of the Chern form plus the Euler characteristic, $\chi(\mathcal{M})$, upon proper use of the Gauss-Bonnet theorem. Defining the evolution of surfaces in terms of a radial foliation of the spacetime, we can write the metric in Gauss normal coordinates as
\begin{align}
    \diff{s^2} = N^2(z)\diff{z^2} + h_{ij}(z,x)\diff{x^i}\diff{x^j}\,,
\end{align}
where the set $\{x^i\}$ denotes codimension-1 boundary coordinates. Then, the Gauss-Bonnet theorem can be written as
\begin{align}
\int_{\mathcal{M}}\diff{^4x}\sqrt{|g|}\,\mathcal{G} = 32\pi^2\chi(\mathcal{M}) + 4\int_{\partial\mathcal{M}}\diff{^3x}\sqrt{|h|}\delta^{i_1i_2i_3}_{j_1j_2j_3} K^{j_1}_{i_1}\left(\frac{1}{2}\mathcal{R}^{j_2j_3}_{i_2i_3} - \frac{1}{3}K^{j_2}_{i_2} K^{j_3}_{i_3} \right)\,.
\end{align}
where $\mathcal{R}^{ij}_{kl}=\mathcal{R}^{ij}_{kl}(h)$ and $K_{ij}=-\tfrac{1}{2N}\partial_z h_{ij}$ are the intrinsic and extrinsic curvatures, respectively.

Although the Gauss-Bonnet term does not modify the field equations, it changes the action and conserved charges nontrivially. Indeed, if
\begin{align}\label{alphasol}
    \alpha = \frac{\ell^2}{4}\,,
\end{align}
the action and variations thereof become finite upon evaluation on asymptotically AdS spacetimes~\cite{Aros:1999id,Aros:1999kt,Olea:2005gb,Miskovic:2009bm}. In the Fefferman-Graham gauge, the latter can be expressed as
\begin{align}
    \diff{s^2} = \frac{\ell^2}{z^2}\left(\diff{z^2} + \bar{g}_{ij}(z,x)\diff{x^i}\diff{x^j} \right)\,,
\end{align}
where $z\to0$ denotes the location of the conformal boundary. In general, $\bar{g}_{ij}$ admits a series expansion in the holographic coordinate as
\begin{align}
    \bar{g}_{ij}(z,x) = g_{(0)ij}(x) + \frac{z}{\ell}g_{(1)ij}(x)+ \frac{z^2}{\ell^2}g_{(2)ij}(x) + \frac{z^3}{\ell^3}g_{(3)ij}(x) + \mathcal{O}(z^4)\,.
\end{align}
In pure gravity, for Einstein spaces, the condition $g_{(1)ij}=0$ is implied by the field equations. Furthermore, even in Einstein-AdS-Maxwell theory, the falloff of the matter sector still implies the vanishing of $g_{(1)}$~\cite{Caldarelli:2016nni}. Then, taking into account Eqs.~\eqref{alphasol} and~\eqref{GB}, the action~\eqref{EMaction} can be written in a MacDowell-Mansouri form~\cite{MacDowell:1977jt} augmented by the Maxwell term, that is,
\begin{align}\label{MMaction}
    I_{\rm EM}^{\rm (ren)} &= \frac{\kappa\ell^2}{16}\int_{\mathcal{M}}\diff{^4x}\sqrt{|g|}\left[\delta^{\mu_1\ldots\mu_4}_{\nu_1\ldots\nu_4}\left(R^{\nu_1\nu_2}_{\mu_1\mu_2} + \frac{1}{\ell^2}\delta^{\nu_1\nu_2}_{\mu_1\mu_2}\right)\left(R^{\nu_3\nu_4}_{\mu_3\mu_4} + \frac{1}{\ell^2}\delta^{\nu_3\nu_4}_{\mu_3\mu_4}\right) - \frac{4}{\kappa\ell^2}F_{\mu\nu}F^{\mu\nu} \right]\,. 
\end{align}

To evaluate this action on-shell, we use the irreducible decomposition of the Riemann tensor in arbitrary $D$ dimensions
\begin{align}
    R^{\mu\nu}_{\lambda\rho} = \frac{1}{D(D-1)}\delta^{\mu\nu}_{\lambda\rho}R + \frac{4}{D-2}\delta^{[\mu}_{[\lambda}H^{\nu]}_{\rho]} + W^{\mu\nu}_{\lambda\rho} \,,
\end{align}
where $H^\mu_\nu = R^\mu_\nu - \tfrac{1}{D}\delta^\mu_\nu R$ is the traceless Ricci tensor and $W^{\mu\nu}_{\lambda\rho}$ denotes the Weyl tensor. The former vanishes identically upon evaluation on Einstein spaces while the latter is traceless in any of their indices. In $D=4$, the equations of motion~\eqref{eom} imply $R=-12/\ell^2$ and $H_{\mu\nu} = (2\kappa)^{-1}T_{\mu\nu}$. Thus, the Weyl tensor can be written in four dimensions on-shell as
\begin{align}\label{Weylonshell}
    W^{\mu\nu}_{\lambda\rho} = R^{\mu\nu}_{\lambda\rho} + \frac{1}{\ell^2}\delta^{\mu\nu}_{\lambda\rho} - \frac{1}{\kappa}\,\delta^{[\mu}_{[\lambda}\,T^{\nu]}_{\rho]}\,,
\end{align}
where the stress-energy tensor of Maxwell fields is defined in Eq.~\eqref{Tmunu}. Then, replacing this expression into the action~\eqref{MMaction}, we find that the on-shell action is
\begin{align}\label{IEonshellcovariant}
    I_{\rm EM}^{\rm (ren)}\Big|_{\rm on-shell} = \frac{\kappa\ell^2}{4}\int_{\mathcal{M}}\diff{^4}x\sqrt{|g|}\left(W^{\mu\nu}_{\lambda\rho}W^{\lambda\rho}_{\mu\nu} - \frac{1}{\kappa\ell^2}F_{\mu\nu}F^{\mu\nu} - \frac{1}{2\kappa^2}T_{\mu\nu}T^{\mu\nu}\right) \,.
\end{align}
Notice that the first two terms of the on-shell action remain invariant under a Weyl rescaling $g_{\mu\nu}\to e^{2\Omega(x)}g_{\mu\nu}$ and $A_\mu\to A_\mu$. However, the last term is not, since $T_{\mu\nu}\to e^{-2\Omega(x)}T_{\mu\nu}$. Even though conformally invariant actions remain finite in asymptotically AdS spacetimes~\cite{Grumiller:2013mxa}, this is also the case for Einstein-AdS-Maxwell due to the falloff of the matter content, similar to what happens in the presence of conformally-coupled scalar fields~\cite{Anastasiou:2022wjq}. This can be seen by noticing that asymptotically AdS electrovacuum solutions of Einstein-Maxwell theory behave as
\begin{subequations}
    \begin{align}\label{falloff}
W^{\mu\nu}_{\lambda\rho}W^{\lambda\rho}_{\mu\nu}&= W^{ij}_{kl}W^{kl}_{ij} + 4W^{iz}_{jz}W^{jz}_{iz} + 4W^{iz}_{jk}W^{jk}_{iz} \sim\mathcal{O}(z^6)\,, \\  F_{\mu\nu}F^{\mu\nu} &= F_{ij}F^{ij} + 2F_{iz}F^{iz} \sim\mathcal{O}(z^4)\,, \\  T_{\mu\nu}T^{\mu\nu}&= T_{ij}T^{ij} + 2T_{iz}T^{iz} + T_{zz}T^{zz} \sim\mathcal{O}(z^8)\,,  
\end{align}
\end{subequations}
as $z\to0$. Thus, the Einstein-Maxwell on-shell action is finite for asymptotically AdS spacetimes. 

The renormalized Euclidean action for the Kerr-Newman-NUT-AdS solution can be computed explicitly by performing the analytic continuation $t\to -i\tau$, $q\to i\hat{q}$, $p\to i\hat{p}$, $n\to -i\hat{n}$, and $a\to -i\hat{a}$. The absence of conical singularities demands $\tau\sim\tau+\beta_\tau$ and $\varphi\sim\varphi+\beta_\varphi$, where $\beta_\varphi:=-\beta_\tau\Omega_+$ (see, e.g., Refs.~\cite{Gibbons:1979nf,Mann:1996bi,Caldarelli:1999xj}). Indeed, the Hawking temperature is related to the period of the Euclidean time through $T_H=\beta_\tau^{-1}$, where $T_H$ is given in Eq.~\eqref{TH}. Then, direct integration yields
\begin{align}\notag
    I_{\rm EM}^{\rm (ren)}&\Big|_{\rm on-shell} = \frac{2\kappa\beta_\tau\beta_\varphi}{ \ell^2\Xi^2\left[r_+^2 - (\hat{a}+\hat{n})^2 \right]\left[r_+^2-(\hat{a}-\hat{n})^2 \right]}\Bigg[\left(r_+^2+\ell^2\right)^2\left(\hat{a}^4-r_+^4\right)+\hat{n}^2\ell^4\left(2r_+^2-\hat{n}^2\right)\\ \notag
    &+2\hat{n}^2\ell^2\left(4r_+^4-5\hat{n}^2r_+^2+2\hat{n}^4-6\hat{a}^2r_+^2+2\hat{a}^2\hat{n}^2-4\hat{a}^4\right) + \hat{n}^2\big(2r_+^6-4\hat{n}^2r_+^4+6\hat{n}^4r_+^2-3\hat{n}^6 \\ \notag 
    &-12\hat{a}^2r_+^4+40\hat{a}^2\hat{n}^2r_+^2-12\hat{a}^2\hat{n}^4-4\hat{a}^4r_+^2+15\hat{a}^4\hat{n}^2\big) + \frac{\ell^2}{4\kappa}\bigg\{4\hat{n}^2\hat{q}^2r_+^2-8\hat{n}\hat{p}\hat{q}r_+^3 \\
    &-r_+^2(3\hat{p}^2+\hat{q}^2)(\hat{a}^2-r_+^2)+\left[\hat{q}^2-\hat{p}^2 \right]\big[\ell^2(\hat{a}^2+\hat{n}^2-r_+^2)-\hat{n}^4-5\hat{a}^2\hat{n}^2\big]\bigg\}\Bigg]\,,
\end{align}
which is finite as shown previously.\\
Thus, the partition function $\mathcal{Z}$ can be obtained directly from this expression to first order in the saddle-point approximation as $\ln\mathcal{Z}\approx-I_E$. Additionally, there exists a relation between the integration constant such that the solution becomes self-dual~\cite{Alonso-Alberca:2000zeh}. We discuss this case and its main properties in the next section.

\section{Self-duality of Euclidean Kerr-Newman-NUT-AdS instanton\label{sec:SD}}

In the Euclidean section, the dyonic Kerr-Newman-NUT-AdS solution enjoys an (anti-)self-dual curve in parameter space where both Maxwell's field strength and the Weyl tensor satisfy
\begin{align}\label{SDdef}
    F_{\mu\nu}=\pm\tilde{F}_{\mu\nu} \;\;\;\;\; \mbox{and} \;\;\;\;\; W_{\mu\nu\lambda\rho} = \pm\tilde{W}_{\mu\nu\lambda\rho}\,,
\end{align}
respectively, where $\tilde{F}_{\mu\nu}=\tfrac{1}{2}\varepsilon_{\mu\nu\lambda\rho}F^{\lambda\rho}$ and $\tilde{W}_{\mu\nu\lambda\rho} = \tfrac{1}{2}\varepsilon_{\mu\nu\alpha\beta}W^{\alpha\beta}_{\lambda\rho}$. This condition is achieved if the following relations on the parameters are met 
\begin{align}\label{selfdualcond}
      \hat{q} = \pm\,\hat{p} \;\;\;\;\; \mbox{and} \;\;\;\;\; m = \pm \frac{\hat{n}}{G}\left(1+\frac{\hat{a}^2-4\hat{n}^2}{\ell^2}\right)\,.
\end{align}

The stress-energy tensor of Maxwell fields in Eq.~\eqref{Tmunu} vanishes identically for (anti-)self-dual configurations, even though the field strength is nontrivial. This can be seen directly from the identity~\cite{Actor:1979in}
\begin{align}\label{TmunuSD}
    T_{\mu\nu} = F_{\mu\lambda}F_{\nu}{}^{\lambda} - \frac{1}{4}g_{\mu\nu}F_{\lambda\rho}F^{\lambda\rho} = \frac{1}{2}\left(F_{\mu\lambda}+\tilde{F}_{\mu\lambda} \right)\left(F_{\nu\rho} - \tilde{F}_{\nu\rho} \right)g^{\lambda\rho}\,.
\end{align}
Thus, (anti-)self-dual Maxwell solutions do not backreact; they resemble stealth configurations of Einstein gravity (non-)minimally coupled to scalar fields~\cite{Ayon-Beato:2004nzi,Ayon-Beato:2005yoq,Hassaine:2006gz}. Then, using the off-shell identity $\tilde{W}_{\mu\nu\lambda\rho}W^{\mu\nu\lambda\rho}=\tilde{R}_{\mu\nu\lambda\rho}R^{\mu\nu\lambda\rho}$ where $\tilde{R}_{\mu\nu\lambda\rho} = \tfrac{1}{2}\varepsilon_{\mu\nu\alpha\beta}R^{\alpha\beta}_{\lambda\rho}$ is the dual Riemann tensor, we find that the Euclidean on-shell action~\eqref{IEonshellcovariant} can be written as the linear combination of the $SO(4)$ and $U(1)$ Chern-Pontryagin indices, that is, 
\begin{align}
    I_E|_{\rm on-shell} = \pm4\pi^2\left(\kappa\ell^2 P_1[\mathcal{M}] - C_2[A]\right)\,.
\end{align}
This is a topologically-nontrivial minimum of the action and it represents the saturated BPS bound of Einstein-AdS-Maxwell theory. For the (anti-)self-dual dyonic Kerr-Newman-NUT-AdS solution in Eq.~\eqref{solution}, we find that these topological terms are explicitly given by
\begin{align}
    P_1[\mathcal{M}] &= \frac{1}{16\pi^2}\int_{\mathcal{M}}\diff{^4x}\sqrt{|g|}\,\tilde{R}^{\mu\nu}_{\lambda\rho}R_{\mu\nu}^{\lambda\rho} = \pm\frac{2\beta_\tau\beta_\varphi(mG)^2\left[2r_+\left(r_+^2+\hat{a}^2\right)\pm \hat{n}\left(3r_+^2+\hat{a}^2-\hat{n}^2\right) \right]}{\pi^2\Xi^2\left[(r_+\pm\hat{n})^2-\hat{a}^2 \right]^3}\,,\\
    C_2[A] &= \frac{1}{16\pi^2}\int_{\mathcal{M}}\diff{^4x}\sqrt{|g|}\,\tilde{F}_{\mu\nu}F^{\mu\nu} = \pm\frac{\beta_{\tau}\beta_\varphi r_+ \hat{p}^2}{2\pi^2\Xi^2\left[(r_+\pm\hat{n})^2-\hat{a}^2 \right]}\,,
\end{align}
where $m$ and $\hat{p}$ satisfy the conditions in Eq.~\eqref{selfdualcond}. This implies that the (anti-)self-dual dyonic Kerr-Newman-NUT-AdS space is a BPS state that preserves only a fraction of the supersymmetries~\cite{Alonso-Alberca:2000zeh}. These terms, in addition to the boundary contributions coming from the Chern-Simons form and the Atiyah-Patodi-Singer $\eta$-invariant, can be used to compute the Dirac index and to see whether this background breaks the chiral symmetry of massless fermionic fields or not (see~\cite{Eguchi:1980jx} and references therein). 

Since the Euclidean on-shell action becomes the sum of two topological terms when evaluated at (anti-)self-dual configurations, it is natural to modify the action~\eqref{MMaction} as
\begin{align}\label{Ibar}
    \bar{I}_{\rm EM}^{\rm (ren)} &= I_{\rm EM}^{\rm (ren)} -  \frac{\kappa\ell^2}{4} \left(\Theta_{\tilde{R}R}\int_{\mathcal{M}}\diff{^4x}\sqrt{|g|}\,\tilde{R}^{\mu\nu}_{\lambda\rho}R_{\mu\nu}^{\lambda\rho} - \frac{\Theta_{\tilde{F}F}}{\kappa\ell^2}\int_{\mathcal{M}}\diff{^4x}\sqrt{|g|}\,\tilde{F}_{\mu\nu}F^{\mu\nu} \right)\,,
\end{align}
with $\Theta_{\tilde{R}R}$ and $\Theta_{\tilde{F}F}$ two dimensionless constants. These two parity-odd topological terms modify neither the bulk dynamics nor the infrared divergences for asymptotically locally AdS spaces. Nevertheless, they do modify the finite part of the action. Indeed, if $\Theta_{\tilde{R}R}=\Theta_{\tilde{F}F}=\pm1$, the Euclidean on-shell action~\eqref{Ibar} can be written as
\begin{align}\notag
    \bar{I}_{\rm EM}^{\rm (ren)}\Big|_{\rm on-shell} &= \frac{\kappa\ell^2}{8}\int_{\mathcal{M}}\diff{^4x}\sqrt{|g|}\Bigg[\left(W^{\mu\nu}_{\lambda\rho}\mp\tilde{W}^{\mu\nu}_{\lambda\rho}\right)\left(W_{\mu\nu}^{\lambda\rho}\mp\tilde{W}_{\mu\nu}^{\lambda\rho} \right)  - \frac{1}{\kappa\ell^2}\left(F_{\mu\nu}\mp\tilde{F}_{\mu\nu}\right)\left(F^{\mu\nu}\mp\tilde{F}^{\mu\nu}\right) \\
    \label{Ibarfinal}
    &-\frac{1}{8\kappa^2}\left(F_{\mu\lambda}+\tilde{F}_{\mu\lambda} \right)\left(F_{\nu\rho} - \tilde{F}_{\nu\rho} \right)\left(F^{\mu\sigma}+\tilde{F}^{\mu\sigma} \right)\left(F^{\nu\tau} - \tilde{F}^{\nu\tau} \right)g^{\lambda\rho}g_{\sigma\tau}\Bigg]\,,
\end{align}
where we have used Eqs.~\eqref{MMaction} and~\eqref{TmunuSD}. Thus, Euclidean on-shell action~\eqref{Ibarfinal} vanishes identically for (anti-)self-dual electrovacuum solutions with negative cosmological constant---see Eq.~\eqref{SDdef}. Then, it becomes evident that adding topological terms of the Pontryagin class enlarges the ground state to all (anti-)self-dual configurations of Einstein-AdS-Maxwell theory, similar to what happens with the gravity and matter sectors separately. 

\section{Conserved charges in AdS gravity and electric/magnetic duality\label{sec:AMD}}

The (anti-)self-duality condition on the Weyl tensor allows one to establish an electric/magnetic duality between the AMD charges of the Kerr-Newman-NUT-AdS solution.\footnote{In the absence of the NUT charge and Maxwell fields, AMD charges of the Kerr-AdS black hole have been computed in Ref.~\cite{Das:2000cu} in different dimensions.} To show this, we consider the electric and magnetic parts of the Weyl tensor defined as 
\begin{align}
    \mathscr{E}_{ij} = W_{i\lambda j\rho}n^\lambda n^\rho \;\;\;\;\; \mbox{and} \;\;\;\;\; \mathscr{B}_{ij} = \tilde{W}_{i\lambda j\rho}n^\lambda n^\rho\,,
\end{align}
respectively, where $n^\mu$ is the unit-normal spacelike vector that defines radial foliation. Since the Maxwell fields decay sufficiently fast towards the asymptotic boundary, its presence does not affect the condition $g_{(1)}=0$ in Einstein gravity (see Ref.~\cite{Caldarelli:2016nni} and references therein). 

To relevant order in the FG expansion, variations of the action~\eqref{MMaction} lead to the holographic part of the renormalized Balasubramanian-Kraus stress-energy tensor $\tau_{ij}$~\cite{Miskovic:2009bm} which, in turn, is related to the electric part of the Weyl tensor via~\cite{Miskovic:2009bm,Bakas:2008gz,Bakas:2009pbm}
\begin{align}
    \tau_{ij} = -\frac{\ell}{8\pi G}\,\mathscr{E}_{ij}\,.
\end{align}
On the other hand, near the conformal boundary, the magnetic part of the Weyl tensor can be expressed to the relevant holographic order as~\cite{Miskovic:2009bm,Bakas:2008gz,Bakas:2009pbm}
    \begin{align}
    -\ell\, C_{ij}(g_{(0)}) = \mathscr{B}_{ij}\,, \;\;\;\;\; \mbox{where} \;\;\;\;\; C^{ij}(g_{(0)}) = \varepsilon^{(0)ilm} \nabla_{(0)m}(R^j_{(0)l}-\frac{1}{4}R_{(0)}\delta^j_l)
\end{align}
is the Cotton tensor and all quantities in the last expression are constructed out of $g_{(0)}$. Explicitly, the components of the holographic stress tensor and Cotton tensor are
\begin{align}\label{stresscottoncomp}
    \tau_{ij} = \frac{m}{8\pi\ell\Xi^2}\,\sigma_{ij}(\vartheta) \;\;\;\;\; \mbox{and} \;\;\;\;\; C_{ij} = \frac{n}{\Xi^2\ell^3}\left(1-\frac{a^2-4n^2}{\ell^2} \right)\sigma_{ij}(\vartheta)\,,
\end{align}
respectively, where we have defined the codimension-1 boundary tensor as
\begin{align}
\sigma_{ij}(\vartheta) = \begin{pmatrix}
   2 & 0 & 4n(C+\cos\vartheta) - 2a\sin^2\vartheta \\
   0  & \frac{\ell^2\Xi^2}{\Delta_\vartheta(\vartheta)} & 0 \\
   4n(C+\cos\vartheta) - 2a\sin^2\vartheta & 0 & \ell^2\sin^2\vartheta\Delta_\vartheta(\vartheta) + 2\left[2n(C+\cos\vartheta) - a\sin^2\vartheta \right]^2
\end{pmatrix} ,
\end{align}
and we have chosen the orientation $(t,\vartheta,\varphi)$. From Eq.~\eqref{stresscottoncomp}, we conclude that the holographic stress tensor can be written in terms of the Cotton tensor as
\begin{align}\label{tauandcotton}
    \tau_{ij} = \frac{m\ell^2}{8\pi n}\left(1-\frac{a^2-4n^2}{\ell^2} \right)^{-1} \, C_{ij}\,.
\end{align}
This relation is known as the stress tensor/Cotton tensor duality and it appears also in the fluid interpretation of the holographic stress tensor~\cite{Leigh:2007wf,Caldarelli:2011idw,Mansi:2008br,Mansi:2008bs,Miskovic:2009bm,Bakas:2008gz,Bakas:2009pbm,Leigh:2011au,Leigh:2012jv,Mukhopadhyay:2013gja}. 

For a Killing vector $\xi=\xi^\mu\partial_\mu$, one can define conserved charges associated with the electric and magnetic parts of the Weyl tensor as long as the matter flux vanishes at the asymptotic boundary~\cite{Ashtekar:1984zz,Ashtekar:1999jx}. Thus, the conserved electric and magnetic charges are given by
\begin{subequations}\label{QAMD}
\begin{align}\label{Qelectric}
    \mathcal{Q}[\xi] &= -\frac{\ell}{8\pi G}\int_{\Sigma_\infty}\diff{^2x}\sqrt{|\gamma|}\;\mathscr{E}^i_j\, \xi^j \, u_i \,, \\
    \label{Qmagnetic}
    \tilde{\mathcal{Q}}[\xi] &= -\frac{\ell}{8\pi G}\int_{\Sigma_\infty}\diff{^2x}\sqrt{|\gamma|}\;
\mathscr{B}^i_j\, \xi^j\, u_i \,,
\end{align}    
\end{subequations}
where $\Sigma_\infty$ is a codimension-2 asymptotic boundary with $u^i$ its timelike unit normal, while $\gamma$ denotes the determinant of the codimension-2 boundary metric. For an asymptotically timelike Killing vector field $\xi=\partial_t$, direct evaluation of these integrals yields
\begin{subequations}\label{AMDMass}
\begin{align}
    \mathcal{Q}[\partial_t] &:= M = \frac{m}{\Xi^2} \, , \\ 
    \tilde{\mathcal{Q}}[\partial_t] &:= \tilde{M} = \frac{n}{G\Xi^2}\left(1 - \frac{a^2-4n^2}{\ell^2} \right)\,.
\end{align}    
\end{subequations}
The (anti-)self-duality condition $\mathscr{E}_{ij}=\pm\mathscr{B}_{ij}$ (cf. Eq.~\eqref{selfdualcond} in Euclidean signature) implies that $M=\tilde{M}$; the latter is usually interpreted as the magnetic mass~\cite{Araneda:2016iiy,Araneda:2018orn}. Indeed, to relevant holographic order, the (anti-)self-duality condition yields 
\begin{align}\label{selfdualtaucotton}
    \tau_{ij} = \pm\frac{\ell^2}{8\pi G} C_{ij}\,.
\end{align}
For the spacelike Killing vector that generates the axial isometry, i.e. $\xi=\partial_\varphi$, the electric and magnetic charges are given by
\begin{subequations}
    \begin{align}
        \mathcal{Q}[\partial_\varphi] &:= J = M(3Cn-a)\,, \\
        \tilde{\mathcal{Q}}[\partial_\varphi] &:= \tilde{J} = \tilde{M}(3Cn-a)\,,
    \end{align}
\end{subequations}
with $M$ and $\tilde{M}$ given in Eq.~\eqref{AMDMass}. Therefore, the (anti-)self-duality of the dyonic Kerr-Newman-NUT-AdS solution is not only translated into an equivalence between the electric and magnetic masses but also provides a duality between the electric and magnetic parts of the angular momentum. Notice that the position of the Misner string parametrized by $C=\pm1,0$ induces a nontrivial contribution to the angular momentum as well. 

The Noether-Wald formalism~\cite{Wald:1993nt,Iyer:1994ys,Wald:1999wa}, on the other hand, provides a covariant framework to compute conserved charges. The latter is based upon the on-shell conservation of the Noether current associated with diffeomorphism invariance. Locally, the Poincaré Lemma allows one to write the Noether current in terms of a Noether prepotential, that is,
\begin{align}
    q^{\mu\nu} = -2\left(E^{\mu\nu}_{\lambda\rho}\nabla^\lambda\xi^\rho + 2\xi^\lambda\nabla^\rho E^{\mu\nu}_{\lambda\rho} \right)\,.
\end{align}
Here, $\xi=\xi^\mu\partial_\mu$ is a Killing vector field and $E^{\mu\nu}_{\lambda\rho}$ denotes the functional derivative of the Lagrangian with respect to the Riemann tensor. For the action~\eqref{MMaction}, it is explicitly given by
\begin{align}
    E^{\mu\nu}_{\lambda\rho} = \frac{\kappa\ell^2}{8}\delta^{\mu\nu\alpha\beta}_{\lambda\rho\gamma\delta}\left(R^{\gamma\delta}_{\alpha\beta} + \frac{1}{\ell^2}\delta^{\gamma\delta}_{\alpha\beta} \right) = \frac{\kappa\ell^2}{2}\left(W^{\mu\nu}_{\lambda\rho} + \frac{1}{\kappa}\delta^{[\mu}_{[\lambda}T^{\nu]}_{\rho]} \right)\,.
\end{align}
In the last equality, we have used the on-shell relation of Eq.~\eqref{Weylonshell}. Conserved charges associated with a Killing vector field $\xi$ can be defined in terms of the Noether prepotential as
\begin{align}\label{NWcharge}
    Q[\xi] = \int_{\Sigma_\infty} q^{\mu\nu}\diff{\Sigma_{\mu\nu}}\,,
\end{align}
where $\Sigma_\infty$ is a codimension-2 asymptotic boundary with $\diff{\Sigma_{\mu\nu}}$ being its area element. Notice that, as shown in Eq.~\eqref{falloff}, the stress-energy tensor falloff is subleading as compared with the Weyl tensor. Therefore, it contributes neither to the mass nor the angular momentum of the solution. 

The conserved charges of the dyonic Kerr-Newman-NUT-AdS black hole associated with the Killing vectors $\partial_t$ and $\partial_\varphi$ can be computed directly from the topologically-renormalized conserved charges in Eq.~\eqref{NWcharge}, giving
\begin{subequations}
\begin{align}
    Q[\partial_t] &:= M = \frac{m}{\Xi^2} \,,\\
    Q[\partial_\varphi] &:= J = \frac{m(3Cn-a)}{\Xi^2}\,, 
\end{align}    
\end{subequations}
respectively. These quantities coincide with those obtained in Sec.~\ref{sec:AMD} and in Ref.~\cite{Rodriguez:2021hks} up-to-the choice of rotation's direction. Notice that the position of the Misner string parametrized by $C=\pm1,0$ contributes nontrivially to the angular momentum as in the AMD charges. Indeed, these values lead to the correct thermodynamic description of the solution in the absence of Maxwell fields and it coincides with Brown-York charges as shown in Ref.~\cite{Rodriguez:2021hks}. 

Finally, as discussed in Sec.~\ref{sec:SD}, the inclusion of the Pontryagin densities to the Einstein-AdS-Maxwell action [cf. Eq.~\eqref{Ibar}] does modify the action and variations thereof without affecting the bulk dynamics. In the case of Noether-Wald charges, this can be seen by taking the functional derivative of the corresponding Lagrangian in Eq.~\eqref{Ibar} with respect to the Riemann tensor, which yields
\begin{align}
    \bar{E}^{\mu\nu}_{\lambda\rho} =  E^{\mu\nu}_{\lambda\rho} \mp \frac{\kappa\ell^2}{2}\tilde{R}^{\mu\nu}_{\lambda\rho} = \frac{\kappa\ell^2}{2}\left(W^{\mu\nu}_{\lambda\rho}\mp\tilde{W}^{\mu\nu}_{\lambda\rho}\right) + \frac{\ell^2}{2}\delta^{[\mu}_{[\lambda}T^{\nu]}_{\rho]}\,.
\end{align}
In the last equation, we used the on-shell relation given by Eq.~\eqref{Weylonshell}. Taking into account the identity~\eqref{TmunuSD}, it is clear that the Noether prepotential vanishes identically when Eq.~\eqref{SDdef} holds. Therefore, the addition of the $SO(4)$ and $U(1)$ Pontryagin densities to the Einstein-AdS-Maxwell action not only fixes self-dual configurations as the ground state at the level of the action but also at the level of conserved charges. Indeed, the same conclusion is implied by the modification generated by the Pontryagin density on the holographic stress tensor, that is~\cite{Miskovic:2009bm,Araneda:2016iiy,Ciambelli:2020qny},
\begin{align}
    \tau_{ij}^\pm = \tau_{ij} \mp \frac{\ell^2}{8\pi G}C_{ij}\,,
\end{align}
where the last piece comes from variations of the Pontryagin density. Thus, since the (anti-)self-duality condition implies that the holographic stress tensor and the Cotton tensor are related through Eq.~\eqref{selfdualtaucotton}, it becomes clear that $\tau_{ij}^\pm$ vanishes identically along such a curve of parameter space.

\section{Discussion\label{sec:discussion}}

In this work, we studied the (anti-)self-dual conditions of Kerr-Newman-NUT-AdS spacetimes. We showed that a BPS bound is saturated along the curve in parameter space given by Eq.~\eqref{selfdualcond}. This condition not only provides an electric/magnetic duality between the mass and the NUT charge but at the level of the angular momentum as well. This is shown by computing explicitly the Noether-Wald and Ashtekar-Magnon-Das charges, alongside the holographic stress tensor and boundary Cotton tensor, rendering the duality of Refs.~\cite{Leigh:2007wf,Caldarelli:2011idw,Mansi:2008br,Mansi:2008bs,Miskovic:2009bm,Bakas:2008gz,Bakas:2009pbm,Leigh:2011au,Leigh:2012jv,Mukhopadhyay:2013gja} manifest.

Using topological renormalization, we computed the Euclidean on-shell action of the Kerr-Newman-NUT-AdS spacetime. At the (anti-)self-dual curve, we showed that it can be written as the linear combination of the $SO(4)$ and $U(1)$ Chern-Pontryagin indices, providing a BPS bound for this five-parameter family of solutions. Thus, adding topological terms of the Pontryagin class, the ground state of the theory is enlarged to all (anti-)self-dual configurations of Einstein-AdS-Maxwell theory. We prove this at the level of the action and conserved charges.

Interesting questions remain open. As a matter of fact, a holographic fluid interpretation of the Taub-NUT-AdS and Kerr-AdS metric was given in Refs.~\cite{Leigh:2007wf,Caldarelli:2011idw,Mansi:2008br}. The dyonic Kerr-Newman-NUT-AdS solution is a natural generalization of these spacetimes and its dual interpretation is certainly relevant for describing charged holographic fluids with vorticity. On the other hand, the thermodynamics of dynamic black holes has been a subject of great interest, in particular, those belonging to the Plebanski-Demianski family. It would be interesting to extend these results in the presence of acceleration. We postpone these and other questions for future work. 

\begin{acknowledgments}
   We thank Giorgos Anastasiou, Felipe Díaz, and Daniel Flores-Alfonso for valuable comments and remarks. This work is supported by Agencia Nacional de Investigación y Desarrollo (ANID) through  Anillo Grant ACT210100 \emph{Holography and its applications to High Energy Physics, Quantum Gravity and Condensed Matter Systems} and FONDECYT Regular Grants No 1210500, 1230112, 1230492, 1231779, 1240043, 1240048 and 1240955. 
\end{acknowledgments}

\bibliography{References}

\end{document}